\documentclass[letter,
               keeplastbox,   
               ]{jacow}
\usepackage{pdfpages,multirow,ragged2e} %
\makeatletter%
	\ifboolexpr{bool{xetex}}
	 {\renewcommand{\Gin@extensions}{.pdf,%
	                    .png,.jpg,.bmp,.pict,.tif,.psd,.mac,.sga,.tga,.gif,%
	                    .eps,.ps,%
	                    }}{}
\makeatother

\ifboolexpr{bool{xetex} or bool{luatex}} 
 {}                                      
 {\usepackage[utf8]{inputenc}}           

\usepackage[USenglish]{babel}

\ifboolexpr{bool{jacowbiblatex}}%
 {%
  \addbibresource{jacow-test.bib}
  \addbibresource{biblatex-examples.bib}
 }{}
\begin{document}

\title{Multiphysics simulations of thermal shock testing of nanofibrous high power targets}

\author{W. Asztalos\thanks{wasztalos@hawk.iit.edu}, Y. Torun, Illinois Institute of Technology \\
S. Bidhar, F. Pellemoine, Fermi National Accelerator Laboratory \\
		P. Rath, Indian Institute of Technology Bhubaneswar
		}

\maketitle

\begin{abstract}
   Increase of primary beam power for neutrino beam-lines leads to a reduced lifespan for production targets. New concepts for robust targets are emerging from the field of High Power Targetry (HPT); one idea being investigated by the HPT R\&D Group at Fermilab is an electrospun nanofiber target. As part of their evaluation, samples with different densities were sent to the HiRadMat facility at CERN for thermal shock tests. The samples with the higher density, irradiated under a high intensity beam pulse, exhibit major damage at the impact site whereas those with the lower density show no apparent damage. The exact cause of this failure was unclear at the time. In this paper, we present the results of multiphysics simulations of the thermal shock experienced by the nanofiber targets that suggest the failure originates from the reduced permeability of the high density sample to air flow. The air present in the porous target expands due to heating from the beam, but is unable to flow freely in the high density sample, resulting in a larger back pressure that blows apart the nanofiber mat. We close with a discussion on how to further validate this hypothesis.
\end{abstract}

\section{INTRODUCTION}
One of the common methods for producing neutrino beams is sending a high energy proton beam through a fixed target. Since studying rare particle processes requires a high neutrino flux, neutrino facilities with higher and higher primary proton beam powers are in demand. Facilities such as Neutrinos at the Main Injector (NuMI) at Fermilab have been operating near the 1 megawatt threshold, with future facilities such as the Long Baseline Neutrino Facility intending to cross this milestone.

The environment created by such powerful proton beams is quite extreme---rapid temperature rise from energy deposition by the high intensity pulsed beam causes thermal stress, and proton irradiation leads to embrittlement, swelling, and defect formation. Increasing the intensity of the primary beam can only exacerbate these problems, and so the challenge lies more with robust target design than accelerating structures. Researchers in the area of High Power Targetry (HPT) study new target concepts to meet these future challenges, and many unique ideas have been proposed.

One such project pursued by the HPT Research and Development (HPT R\&D) Group at Fermilab is an electrospun nanofiber target concept, made of ceramics such as Yttria Stabilized Zirconia (YSZ) or tungsten \cite{productionQualification,nanofiberPoster}. The potential advantages are that the solid phase consists of individual fibers that can move freely, which could dissipate thermal stress waves, and that the fibers show resistance to radiation damage without any treatment \cite{productionQualification}. The radiation resistance of the nanofibers is presumed to be due to a small crystal grain size which provides ample defect sinks in the form of grain boundaries. The porosity of the targets also allows for forced gas cooling, with a large exposed surface area.

To evaluate the potential of these targets, the HPT R\&D Group prepared two samples of nanofiber mats and sent them to the HiRadMat Facility at CERN \cite{hrm1,hrm2} for prototypic thermal shock testing in 2018, a way to assess target survivability in an intense pulsed beam environment. The samples were both made of YSZ, with the difference being how tightly packed the fibers were, which we quantify using the Solid Volume Fraction (SVF), $f$, of the samples. The SVF is the fraction of the total volume occupied by \textit{solid} material; based on measurements in \cite{productionQualification}, the low density sample had ${f \approx 0.05}$, whereas the high density sample had ${f \approx 0.20}$.

\begin{figure}[!htb]
   \centering
   \includegraphics*[width=.9\columnwidth]{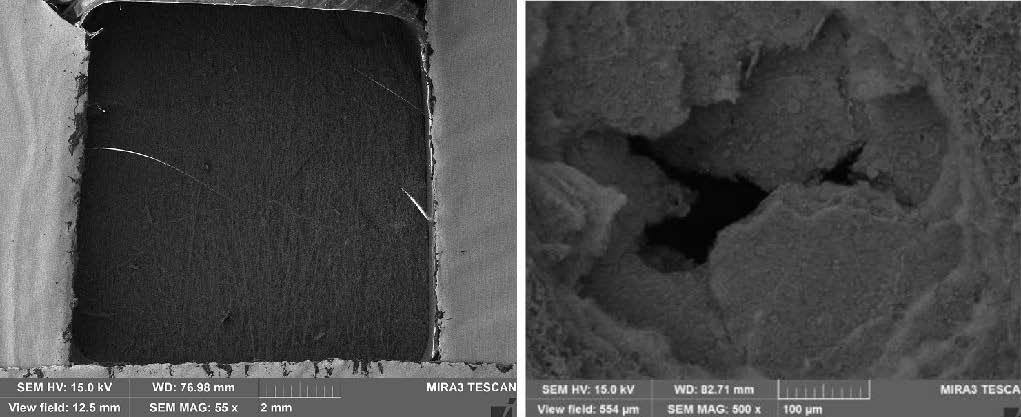}
   \caption{SEM images from HiRadMat test post-irradiation. Left: $f=0.05$ Right: $f=0.20$. Courtesy of Sujit Bidhar.}
   \label{fig:hrm}
\end{figure}

The outcome of the testing was that the low density sample showed no damage from the beam exposure, whereas the high density sample developed a hole at its center, as shown in Fig.~\ref{fig:hrm}. This demonstrates that the survivability of nanofiber targets depends on their construction parameters, and so understanding the mode of failure is critical to the nanofiber target concept's success. In this paper, we share the results of multiphysics simulations of the beam heating of nanofiber mats to better understand the cause of such failures. One of our hypotheses to describe the target failure is that the high density sample failed due to pressure formed by the expansion of air inside the pores which was unable to escape because of the higher viscous resistance.

\section{Model}
To set up these simulations, we require a mathematical model of the nanofiber targets' thermal and fluid dynamical properties, and of the energy deposition by the HiRadMat beam. There are many challenges with modeling nanofibrous structures, as sizing effects give individual fibers properties potentially very different than the bulk materials'. Additionally, simulating individual nanofibers is only possible if we study a very small portion of the target, since their diameters are on the order of hundreds of nanometers, while the RMS beam size of the HiRadMat beam is about ${\sigma=\SI{0.25}{mm}}$. To study large-scale behavior like the response of an entire sample to a beam pulse, we must move to the mesoscale, where we attempt to include the nanoscale effects. To accomplish this, we used Porous Media Models (PMMs) and effective material properties.

In particular, we first used models shared in \cite{insulation} which can be used to calculate the thermal conductivity of a gas inside a porous medium and the thermal conductivity of a fibrous network's solid phase. To get the effective thermal conductivity of the entire target, we used a nonlinear combination of the solid and gas phase thermal conductivities developed by Bhattacharyya \cite{Bhattacharyya}. The heat capacity for each phase is taken to be the same as the bulk materials'.

To handle the flow of air through the porous nanofiber target, we used Darcy's Law, a well-known PMM in the field of fluid dynamics \cite{darcy}. This is a volume-averaged version of the usual Navier-Stokes equations for fluid flow, and the net effect is to introduce a momentum source term to the governing equation. Based on experimental evidence for nanofibrous materials \cite{C2}, and since the Reynolds number is estimated to be small, viscous effects should dominate inertial effects, meaning the source takes the simple form:
\begin{equation}
\vec{S}= - \dfrac{\mu}{\alpha} \; \vec{u}
\end{equation}
where $\mu$ is the dynamic viscosity of the fluid phase, $\alpha$ is the permeability of the porous structure to fluid flow, and $\vec{u}$ is the velocity field. Thus, characterizing the fluid flow in the target just requires an estimate of the permeability, $\alpha$. Our nanofiber targets are electrospun, and thus can be approximated as a layered collection of 2D planes where the fibers are oriented randomly in each individual plane. As such, we used a model for such structures shared in \cite{permeability}:
\begin{equation} \label{eq:perm}
  \alpha = \dfrac{\epsilon \; r^2}{8 \left( \ln \epsilon \right)^2} \; \dfrac{\left( \epsilon - \epsilon_p \right)^{x+2}}{\left( 1-\epsilon_p \right)^x \; \left[ \left( x+1 \right) \epsilon - \epsilon_p \right]^2}  
\end{equation}
where $\epsilon$ is the \textit{porosity} of the sample (related to the SVF by ${\epsilon=1-f}$), $r$ is the average nanofiber radius, \SI{156.5}{nm} in our case, ${\epsilon_p=0.11}$, and $x$ is a parameter which depends on the direction of fluid flow. For the direction along the beam, which is perpendicular to the layering of the fibers, ${x=0.785}$, whereas for the directions inside the planes of fibers, ${x=0.521}$. 

\section{Simulations}
To recreate the HiRadMat experiment, we then applied this model to two YSZ nanofiber targets which were identical except for their SVFs: the low density case with ${f=0.05}$, and the high density case with ${f=0.20}$. We used the MARS program \cite{MARS1,MARS2,MARS3} to calculate the energy deposition in the target due to the beam. MARS cannot model porous or nanoscale structures, and so the difference between the simulations for each sample is seen in the density of the material used. We matched the MARS beam parameters to the HiRadMat facility, so we used a gaussian beam of \SI{440}{GeV} protons with ${\sigma=\SI{0.25}{mm}}$ and ${1.21 \times 10^{13}}$ protons on target, over a pulse lasting \SI{4}{\textmu s}.

The simulation geometry, drawn in Fig~\ref{fig:domain} below, included a \SI{10}{mm} $\times$ \SI{10}{mm} $\times$ \SI{0.1}{mm} rectangle, representing the target, with two \SI{10}{mm} $\times$ \SI{10}{mm} $\times$ \SI{5}{mm} columns of air in front of and behind the target, from the perspective of the beam (the $\hat{z}$ direction). Homogeneous Dirichlet velocity conditions and Neumann heat flux conditions were applied to the walls in the $\pm\hat{x}$ and $\pm\hat{y}$ directions, and homogeneous Dirichlet pressure conditions with a constant temperature of \SI{300}{K} were set at the $\pm \hat{z}$ outlets. The regions were discretized with \SI{0.02}{mm} cubes in the porous zone and \SI{0.1}{mm} cubes in the free columns of air, and the total simulation time was \SI{40}{\textmu s}, with a timestep of \SI{0.1}{\textmu s} during the first \SI{4}{\textmu s} (the beam pulse), and a timestep of \SI{0.5}{\textmu s} after that.

\begin{figure}[!htb]
   \centering
   \includegraphics*[width=.9\columnwidth]{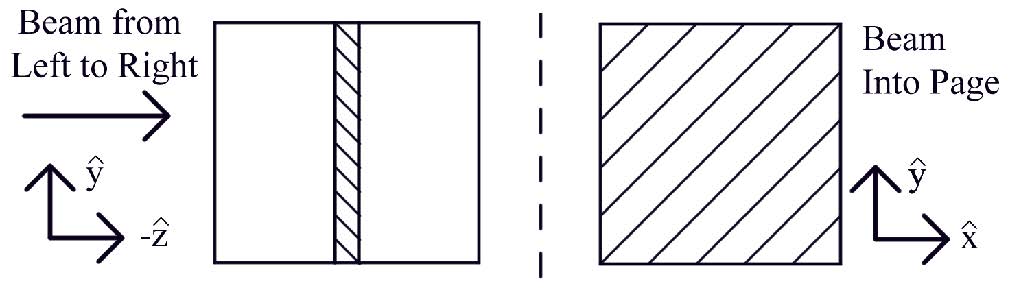}
   \caption{Illustration of problem domain (not to scale). Hatching indicates the nanofiber porous zone.}
   \label{fig:domain}
\end{figure}

For our solver, we chose ANSYS Fluent \cite{fluent}, and used the SIMPLE algorithm to solve for the transient flowfield in the laminar regime, with the energy equation solved simultaneously, coupled to the momentum equation with a pressure and temperature dependent density. We used the equilibrium porous media model in Fluent, specifying fluid permeability according to Eq.~\ref{eq:perm}. The nonequlibrium behavior is handled by the effective material properties. The beam was treated as a volumetric heat source in the porous zone fitted to the MARS results, with a Gaussian dependence on the radial distance from the target center.

The first two simulations compared the ${f=0.05}$ and ${f=0.20}$ samples at standard operating conditions at HiRadMat: an ambient pressure of \SI{1}{atm} and temperature of \SI{300}{K}. To further validate our hypothesis that the failure of the ${f=0.20}$ sample was due to trapped air expanding, we ran 6 additional simulations, where the operating pressure was decreased for the two samples: first \SI{0.1}{atm}, then \SI{0.01}{atm}, and finally \SI{0.001}{atm}, to try and predict the effect of doing the same test with the samples in a vacuum.

\section{Results}
\begin{figure}[!htb]
   \centering
   \includegraphics*[width=.9\columnwidth]{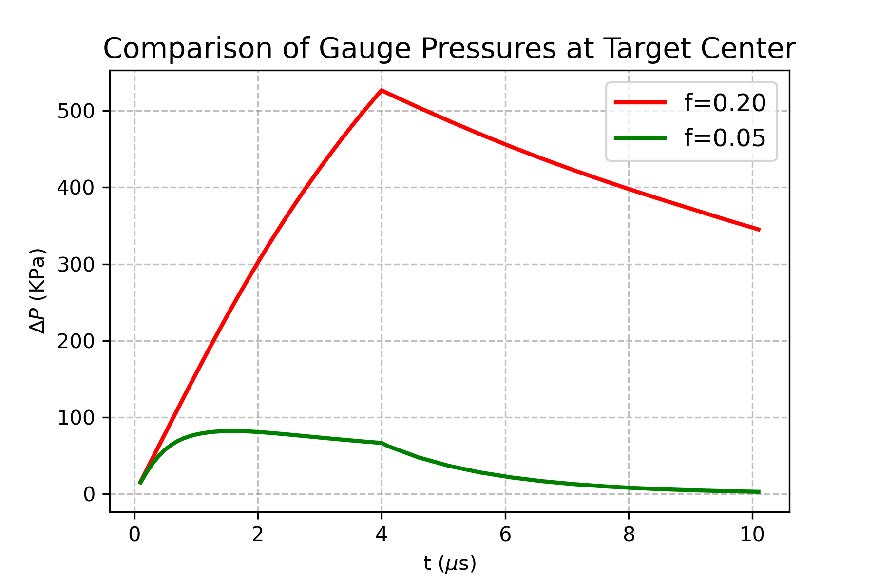}
   \caption{Plot of the pressure rise at the target center for each case at an operating pressure of 1 atmosphere.}
   \label{fig:centerpress}
\end{figure}

The Fluent simulations predicted that the maximum temperature rise experienced by the air inside the mat was ${\Delta T \approx \SI{1880}{K}}$ for ${f=0.05}$ and $\SI{1873}{K}$ for ${f=0.20}$. The increase in energy deposition by the beam for the high density sample was roughly balanced by the change in density and heat capacity, leading to a similar $\Delta T$ for both cases. Figure~\ref{fig:centerpress} shows the gauge pressure, $\Delta P$, computed at the center of the target for each of the two cases at the usual operating pressure of \SI{1}{atm}. Observe that the pressure curve for the high density case, ${f=0.20}$, completely dominates that of the low density case, ${f=0.05}$.

We can also compare the gauge pressures at the boundary of the target, \SI{5}{mm} from the center. The results are shown in Fig.~\ref{fig:boundarypress} below. We can see a pressure wave has propagated from the target center, and that it arrives later in the ${f=0.20}$ case. Both cases have a smaller $\Delta P$ at the boundary, but the ${f=0.20}$ is the smaller of the two now, implying attenuation.
\begin{figure}[!htb]
   \centering
   \includegraphics*[width=.9\columnwidth]{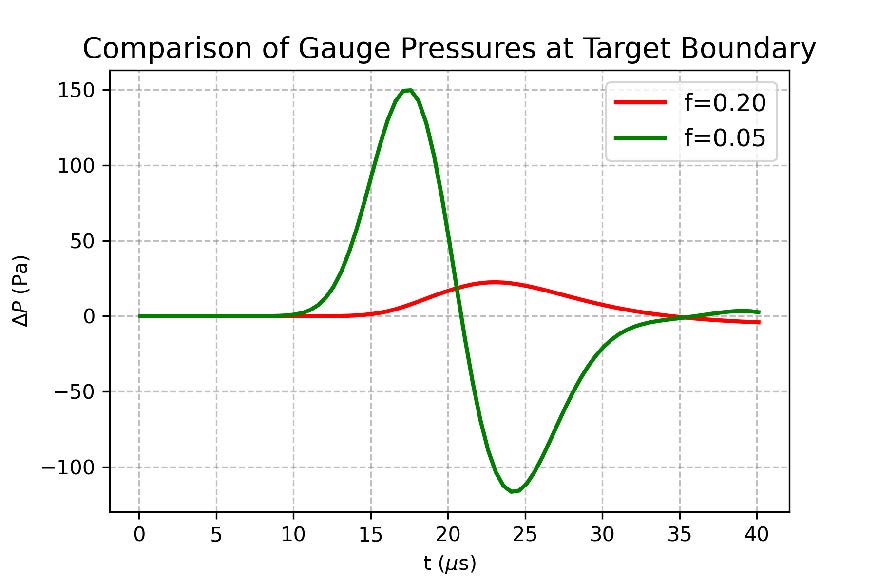}
   \caption{Plot of the pressure rise at the sample boundary for each case at an operating pressure of 1 atmosphere.}
   \label{fig:boundarypress}
\end{figure}

Finally, we examine the effect of lowering the operating pressures and running the same comparisons. In Table~\ref{tab:press}, the gauge pressure calculated at the target center at the end of the beam pulse ($t=\SI{4}{\mu s}$) is listed for both target densities at the four different operating pressures considered. Observe that at the standard operating conditions, the high density sample experiences a much larger $\Delta P$, but the difference vanishes at lower operating pressure.

\begin{table}[!hbt]
   \centering
   \caption{Pressure at Target Center at t=\SI{4}{\textmu s}}
   \begin{tabular}{lcc}
       \toprule
       \textbf{Operating Pressure} & \textbf{f = 0.20}                      & \textbf{f = 0.05} \\
       \midrule
           \SI{1}{atm}         & \SI{527}{kPa}            & \SI{66.6}{kPa}        \\ 
          \SI{0.1}{atm}       & \SI{63.6}{kPa}            & \SI{42.1}{kPa}        \\ 
           \SI{0.01}{atm}        & \SI{6.33}{kPa}            & \SI{6.42}{kPa}        \\ 
           \SI{0.001}{atm}       & \SI{0.629}{kPa}            & \SI{0.639}{kPa}        \\
       \bottomrule
   \end{tabular}
   \label{tab:press}
\end{table}

\section{Discussion}
These results support our hypothesis that the failure of the high density sample at HiRadMat was due to pressure formed by the expansion of air inside of the pores of the sample in response to heating by the beam; the pressure rise was more than 5 times higher for the ${f=0.20}$ sample at the usual operating pressure of \SI{1}{atm}. Furthermore, as the operating pressure fell, the pressure rise in the target decreased by orders of magnitude as well, and became very similar for the two samples, as shown in Table~1. This suggests that if the test were conducted in a vacuum, the high density sample may not have blown apart.

The actual reason for the difference in behavior lies in the permeabilities. The permeability to fluid flow, $\alpha$, of the ${f=0.05}$ sample was significantly higher than the ${f=0.20}$ sample; for flow in the direction of the beam, ${\alpha \sim 3\times 10^{-13}}$~\SI{}{m}$^2$ for ${f=0.05}$, whereas ${\alpha \sim 10^{-14}}$~\SI{}{m}$^2$ for ${f=0.20}$. This means that when the heated air expands, it encounters more resistance in the high density sample and thus becomes more pressurized, eventually having enough force to blow apart the nanofibrous structure. This resistance can also be seen in Fig.~\ref{fig:boundarypress}, as the pressure wave from the beam pulse takes longer to reach the boundary in the ${f=0.20}$ case, and is more attenuated.

\section{CONCLUSION}
These multiphysics simulations support our hypothesis that the failure mechanism of the high density nanofiber target sample at the HiRadMat test was due to the air in the pores of the sample encountering more resistance to flow. This work also suggests that running a test under the same beam parameters but with the samples in vacuum should result in both targets surviving, a hypothesis we hope to confirm during a future experiment at HiRadMat organized by the HPT R\&D Group. In the future, we also plan to run more of these ``re-enactment'' simulations to compare with results from another experiment run at HiRadMat in 2022, and to also try and predict the outcomes of the next HiRadMat test using this simulation infrastructure.

\section{ACKNOWLEDGMENTS}
This work was produced by Fermi Research Alliance, LLC under Contract No. DE-AC02-07CH11359 with the U.S. Department of Energy, Office of Science, Office of High Energy Physics. The research presented here was possible with the support of the Fermilab Accelerator PhD Program.

%
%
\ifboolexpr{bool{jacowbiblatex}}%
	{\printbibliography}%

\begin{thebibliography}{9} 

    \bibitem{productionQualification}
        S. Bidhar \textit{et al.}, \textquotedblleft{Production and qualification of an electrospun ceramic nanofiber material as a candidate future high power target}\textquotedblright, \textit{Phys. Rev. Accel. Beams}, vol. 24, p. 123001, Dec 2021.\\ \url{doi:10.1103/PhysRevAccelBeams.24.123001}

    \bibitem{nanofiberPoster}
        S. Bidhar, \textquotedblleft{Electrospun nanofiber materials for high power target applications}\textquotedblright, Fermi National Accelerator Lab, Batavia, IL, USA, 2017, Tech. Rep..

	\bibitem{hrm1}
   		F. J. Harden, A. Bouvard, N. Charitonidis, and Y. Kadi, \textquotedblleft{HiRadMat: A Facility Beyond the Realms of Materials Testing}\textquotedblright, in \textit{Proc. IPAC’19}, Melbourne, Australia, May 2019, pp. 4016--4019.\\
   		\url{doi:10.18429/JACoW-IPAC2019-THPRB085}   


	\bibitem{hrm2}
  		I. Efthymiopoulos \textit{et al.}, \textquotedblleft{HiRadMat: A New Irradiation Facility for Material Testing at CERN}\textquotedblright,
     	in \textit{Proc. IPAC’11}, San Sebastian, Spain, Sep. 2011, paper TUPS058, pp. 1665--1667.   

    
 \bibitem{insulation}
        S. Carvajal \textit{et al.}, \textquotedblleft{Comparison of models for heat transfer in high-density fibrous
insulation}\textquotedblright, \textit{J. Res. Natl. Inst. Stand. Technol.}, pp. 1–21, 2019. \\
\url{doi:10.6028/jres.124.010}
    
     
     \bibitem{Bhattacharyya}
        R. Bhattacharyya, \textquotedblleft{Heat-transfer model for fibrous insulations}\textquotedblright, \textit{ASTM special
technical publications}, pp. 272–286, 1980. \\
\url{doi:10.1520/STP29279S}


   \bibitem{darcy}
        S. Whitaker, \textquotedblleft{Flow in porous media I: A theoretical derivation of Darcy’s law}\textquotedblright,
\textit{Transp. Porous Med.}, p. 3–25, 1986. \\
\url{doi:10.1007/BF01036523}


    \bibitem{C2}
        O. Akampumuza \textit{et al.}, \textquotedblleft{Analyzing the effect of nanofiber orientation on membrane
filtration properties with the progressive increase in its thickness: a numerical
and experimental approach}\textquotedblright, \textit{Textile Research Journal}, vol. 90, no. 1,
pp. 24–36, 2020. \\
\url{doi:10.1177/0040517519855316}
    

     \bibitem{permeability}
        M. Tomadakis and T. Robertson, \textquotedblleft{Viscous permeability of random fiber structures:
Comparison of electrical and diffusional estimates with experimental and
analytical results}\textquotedblright, \textit{Journal of Composite Materials}, vol. 39, no. 2, pp. 163–188,
2005. \\
\url{doi:10.1177/0021998305046438}


    \bibitem{MARS1}
        N.V. Mokhov and C.C. James, \textit{The MARS Code System User’s Guide, Version
15 (2016)}. Fermilab-FN-1058-APC, 2017. \\
\url{doi:10.2172/1462233}

    \bibitem{MARS2}
        N. Mokhov \textit{et al.}, \textquotedblleft{MARS15 code developments driven by the intensity frontier
needs}\textquotedblright, \textit{Prog. Nucl. Sci. Technol.}, pp. 496–501, 2014. \\
\url{doi:10.15669/pnst.4.496}

    \bibitem{MARS3}
        N. Mokhov, \textquotedblleft{Status of MARS code}\textquotedblright, Fermilab-Conf-03/053, (Batavia, Illinois),
2003.


 
    \bibitem{fluent}
    ANSYS, Inc., \textquotedblleft{Fluent theory guide 15th edition}\textquotedblright, 2013.
	\end{thebibliography}
	{%
	


} 

%
%


\end{document}